\documentclass[fleqn,usenatbib]{mnras}

\usepackage{newtxtext,newtxmath}

\usepackage[T1]{fontenc}

\DeclareRobustCommand{\VAN}[3]{#2}
\let\VANthebibliography\thebibliography
\def\thebibliography{\DeclareRobustCommand{\VAN}[3]{##3}\VANthebibliography}

\usepackage{xcolor}
\usepackage{color}

\usepackage{graphicx}	
\usepackage{amsmath}	

\title[Circumbinary planets around subdwarf binaries]{Combining \texttt{REBOUND} and \texttt{MESA}: Dynamical Evolution of Planets Orbiting Interacting Binaries}

\author[Xing et al.]
{
Zepei Xing$^{1,2}$\thanks{E-mail: zepei.xing@unige.ch },
Santiago Torres$^{3}$\thanks{E-mail: santiago.torres@ist.ac.at},
Ylva G\"otberg$^{3}$,  
Alessandro A. Trani$^{4}$,  
Valeriya Korol$^{5}$, 
\newauthor
and Jorge Cuadra$^{6}$
\\
$^{1}$ Departement d’ Astronomie, Université de Genève,Chemin Pegasi 51,CH-1290 Versoix, Switzerland\\
$^{2}$ Gravitational Wave Science Center (GWSC), Université de Genève, CH1211 Geneva, Switzerland\\
$^{3}$ Institute of Science and Technology Austria (ISTA), Am Campus 1, 3400 Klosterneuburg, Austria \\
$^{4}$ Niels Bohr International Academy, Niels Bohr Institute, Blegdamsvej 17, 2100 Copenhagen, Denmark \\
$^{5}$ Max Planck Institute for Astrophysics, D-85741 Garching, Germany \\
$^{6}$ Departamento de Ciencias, Facultad de Artes Liberales, Universidad Adolfo Ibáñez, Av. Padre Hurtado 750, Viña del Mar, Chile 
}

\date{Accepted XXX. Received YYY; in original form ZZZ}

\pubyear{2024}

\begin{document}
\label{firstpage}
\pagerange{\pageref{firstpage}--\pageref{lastpage}}
\maketitle


\begin{abstract}

Although planets have been found orbiting binary systems, whether they can survive binary interactions is debated. While the tightest-orbit binaries should host the most dynamically stable and long-lived circumbinary planetary systems, they are also the systems that are expected to experience mass transfer, common envelope evolution, or stellar mergers. In this study, we explore the effect of stable non-conservative mass transfer on the dynamical evolution of circumbinary planets. We present a new script that seamlessly integrates binary evolution data from the 1D binary stellar evolution code \texttt{MESA} into the N-body simulation code \texttt{REBOUND}. This integration framework enables a comprehensive examination of the dynamical evolution of circumbinary planets orbiting mass-transferring binaries, while simultaneously accounting for the detailed stellar structure evolution. In addition, we introduce a recalibration method to mitigate numerical errors from updates of binary properties during the system's dynamical evolution. We construct a reference binary model in which a $2.21\, M_\odot$ star loses its hydrogen-rich envelope through non-conservative mass transfer to the $1.76\, M_\odot$ companion star, creating a $0.38\,M_\odot$ subdwarf. 
We find the tightest stable orbital separation for circumbinary planets to be $\simeq 2.5$ times the binary separation after mass transfer. Accounting for tides by using the interior stellar structure, we find that tidal effects become apparent after the rapid mass transfer phase and start to fade away during the latter stage of the slow mass transfer phase. 
Our research provides a new framework for exploring circumbinary planet dynamics in interacting binary systems.

\end{abstract}

\begin{keywords}
orbital dynamics, circumbinary planets, binary stars, subdwarfs 
\end{keywords}

\section{Introduction}
\label{intro}

The existence of circumbinary planets \citep[CBPs, also known as P-type systems,][]{1984CeMec..34..369D} offers valuable insights into the underlying physics involved in planet formation and the dynamical evolution of planetary systems. After the discovery of the first CBP Kepler-16 b \citep{2011Sci...333.1602D}, a series of CBPs have been reported from the \texttt{Kepler} \citep{2010Sci...327..977B} and \texttt{TESS} missions \citep{2015JATIS...1a4003R}. Currently, more than $20$ binary systems have been confirmed to host circumbinary planets, according to the Extrasolar Planets Encyclopaedia\footnote{\href{https://exoplanetarchive.ipac.caltech.edu}{https://exoplanetarchive.ipac.caltech.edu}} and NASA Exoplanet Archive\footnote{\href{https://exoplanet.eu/planets_binary_circum/}{https://exoplanet.eu/planets\_binary\_circum/}}. About 10 of them are orbiting binary systems that contain an evolved star, such as a white dwarf (WD) or a subdwarf. The recently suggested existence of a possible hot Jupiter around the subdwarf and M-dwarf binary Kepler 451 \citep{2022MNRAS.511.5207E} challenges further our understanding of planet formation and dynamics in the context of binary interactions. These binaries are suggested to be post-common envelope binaries \citep[PCEBs,][]{2013A&A...549A..95Z} and are implied to have experienced a dramatic mass transfer episode and a subsequent unstable mass transfer triggering a common envelope (CE) phase. During CE evolution, binaries undergo a rapid and significant orbital shrinkage 
\citep{2013A&ARv..21...59I}, ultimately leading to either a merger or the ejection of the CE. It remains unclear whether circumbinary planets form before or after the CE phase. Some studies suggest that these circumbinary planets around PCEBs are the second-generation planets formed from the ejecta of CE \citep{2013A&A...549A..95Z,2014A&A...563A..61S}. However, \citet{2014MNRAS.444.1698B} argued that, in certain populations, the circumbinary planets are more likely to be the first-generation planets formed prior to CE evolution.

Subdwarfs are low-mass ($\sim 0.35-1\,M_{\odot}$) exposed helium cores that result from envelope-stripping through CE ejections \citep[e.g.][]{2022A&A...666A.182S}, or stable mass transfer \citep{2003MNRAS.341..669H, 2017A&A...605A.109V}. Thousands of subdwarfs are known \citep{2020A&A...635A.193G}, and more and more of their binary companions are being discovered and characterized. The binary companions include low-mass main-sequence stars, WDs, and even more massive Be stars \citep[e.g.][]{2015A&A...576A..44K,2022A&A...666A.182S,2021AJ....161..248W,2024ApJ...962...70K}. The significant orbital evolution and potentially substantial mass loss that are associated with the formation of a subdwarf challenges the existence of circumbinary planets, but could also provide opportunities for new planetary system architectures to develop. Given the confirmed existence of planets orbiting a tight binary containing a WD \citep[e.g.][]{2023MNRAS.523.5086R}, it is clear that binary interactions, although violent and dynamic, do not prohibit the presence of planetary systems. 
Therefore, understanding the impact of subdwarf formation on surrounding planetary systems constitutes one of the most promising avenues for revealing how binary evolution affects planetary systems in general.

Apart from CE evolution, stable mass transfer is also an essential process for the formation of helium WDs \citep[e.g.][]{2018ApJ...858...14S,2020ApJ...889...49B} and subdwarfs \citep{2018A&A...615A..78G} in binary systems. The stable mass transfer process can be modeled using 1D stellar evolution codes \citep{2008MNRAS.384.1109E,2015ApJS..220...15P}, bolstered by a more comprehensive understanding compared to the CE process. Since our understanding of envelope-stripping through CE ejection is still associated with major uncertainties, the impact the planetary system suffers as a result of the ejection is hard to determine. 
Because of these large uncertainties, earlier studies on the dynamical response of circumbinary planets in CE evolution have adopted simplified approximations, including assuming a constant mass loss rate for the binary system during the CE phase and ignoring the mechanical impact of the ejecta \citep{2013MNRAS.429L..45P,2016ApJ...832..183K}. However, planetary systems orbiting mass-transferring binaries can be treated with a more detailed and accurate approach. Because the envelope-stripping mechanism is better understood and takes substantially longer than the CE evolution. In light of this, we investigate the dynamical stability of planetary systems around binaries that undergo stable mass transfer, integrating detailed 1D stellar evolution simulations. 

Given the advancements in ongoing and forthcoming surveys (e.g., Roman \citep{2015arXiv150303757S,2019ApJS..241....3P,2020AJ....160..123J} and TESS \citep{2016SPIE.9904E..2BR}) that seek planetary systems across diverse host systems, we expect to discover an escalating number of CBPs around various binary systems. These discoveries and insights into expected system architectures and planetary conditions, including stability and habitability \citep{2019AstL...45..620S}, represent an exciting new direction for planetary science. To prepare for these anticipated findings, further theoretical and numerical explorations of planetary dynamics in conjunction with interacting binaries will be essential. 

In this work, we combine N-body simulations with detailed stellar and binary evolution models to explore the dynamical evolution of circumbinary planets around binaries through a stable mass transfer phase that leads to the production of a subdwarf B (sdB) binary. In Section~\ref{sec2}, we introduce the {\it N-Body Binary Stellar Evolution} (\texttt{NBSE}) tool, which couples the stellar evolution code Modules for Experiments in Stellar Astrophysics \citep[\texttt{MESA},][]{2011ApJS..192....3P,2013ApJS..208....4P,2015ApJS..220...15P,2018ApJS..234...34P,2019ApJS..243...10P, 2023ApJS..265...15J} and the N-body code \texttt{REBOUND} \citep{2012A&A...537A.128R,2018MNRAS.473.3351R,2020MNRAS.491.2885T,2022MNRAS.510.6001B}. In Section~\ref{sec3}, we demonstrate the application of \texttt{NBSE} to the dynamical evolution of a single circumbinary planet around an interacting binary star. The impact and the implementation of tidal effects due to the interaction of the binary system and CBPs are discussed in Section~\ref{sec4}. Finally, we summarize in Section~\ref{sec5}.

\section{N-Body Binary Stellar Evolution (\texttt{NBSE})}
\label{sec2}

The dynamical evolution of planets around single-evolved stars has been widely studied \citep[e.g.,][] {1996ApJ...470.1187R,2007ApJ...661.1192V,2009ApJ...705L..81V,2012ApJ...761..121M,2013MNRAS.431.1686V,2014MNRAS.437.1404M,2016RSOS....350571V,2016MNRAS.458.3942V,2018A&A...618A..18R,2020ApJ...898L..23R,2024arXiv240509399M}. These studies include simple stellar evolution models, mainly using the stellar tracks in the single stellar population synthesis code \texttt{SSE} \citep{2000MNRAS.315..543H}. For multiple stellar systems, \citet{2021MNRAS.502.4479H} presented the population synthesis code Multiple Stellar Evolution that includes planets, based on the binary population synthesis code \texttt{BSE} \citep{2002MNRAS.329..897H} with simplified binary evolution treatments.

Recently, \citet{2022MNRAS.510.6001B} introduced a machine-independent implementation of parameter interpolation and a constant time-lag model for tides without evolving spins in \texttt{REBOUNDx} \citep{2020MNRAS.491.2885T}. This approach allows results from other integration codes to be used as input parameters for \texttt{REBOUND}. As an example of their technique, they integrated stellar evolution data for single stars from \texttt{MESA} into \texttt{REBOUND}, using their interpolation scheme to update stellar parameters such as mass and radius as a function of time. They demonstrated this by simulating the Sun’s post-main-sequence influence on the outer giant planets.

In this work, we focus on accurately simulating the binary evolution, particularly the mass transfer phase, together with a circumbinary planetary system.  Similar to \citet{2022MNRAS.510.6001B}, we use the state-of-the-art, open-source stellar evolution code \texttt{MESA} together with the high-performance N-body simulation code \texttt{REBOUND} to study the dynamical evolution of CBPs. Compared to \texttt{BSE}-like codes, where the mass transfer rate is obtained based on parametric methods, \texttt{MESA} enables us to calculate mass transfer rates self-consistently considering the rotation of the stars and account for mass and angular momentum loss both through stellar winds and non-conservative mass transfer and tidal effects all along the binary evolution. In this study, we built \texttt{NBSE}\footnote{\texttt{NBSE} is made publicly available upon publication.}, an integrated tool coupling \texttt{MESA} and \texttt{REBOUND}, serving  for the study of the dynamical evolution of CBPs involving not only the accurate calculation of single stars but also the binary evolution.


\subsection{MESA Binary Model}
\label{sec2.1}

First, we construct a binary model that undergoes stable mass transfer, leading to envelope stripping and the formation of an sdB star binary. The reference binary model we compute consists a primary star with $M_{1} = 2.21\,M_{\odot}$, a secondary star with $M_{2} = 1.76\,M_{\odot}$, and an initial orbital period of $6$ days in a circular orbit. It represents one of the sdB-forming binaries at the low-mass end of the stripped star binary grids in \citet{2018A&A...615A..78G}. These initial parameters of the binary stars are chosen to ensure that planets have enough time to form around the central binary. Consequently, we focus on lower initial masses. 

To construct the binary evolution model, we adopt the \texttt{MESA} setup for constructing binary grids within the \texttt{POSYDON} binary population synthesis code \citep{2023ApJS..264...45F}. In this configuration, the \texttt{MESA} \texttt{Dutch} scheme is used for the stellar wind prescription with modifications related to stellar state and surface temperature \citep[see Sec.~3.2.2 of][]{2023ApJS..264...45F}. The tidal effect is treated following the linear approach by calculating the synchronization timescale, distinguishing radiative and convective layers \citep{1981A&A....99..126H,2002MNRAS.329..897H,2018A&A...616A..28Q}. 

To calculate the mass transfer rate from Roche-lobe overflow, the \texttt{Kolb} scheme \citep{1990A&A...236..385K} within \texttt{MESA} is used when the donor star has left the main sequence. In the binary models, mass transfer is generally highly non-conservative due to stellar rotation, meaning that most of the transferred mass is lost from the system. The specific angular momentum of the transferred material follows the implementation of \citet{2013ApJ...764..166D}. During mass transfer, the accretor stars are expected to easily spin up to critical rotation due to accretion \citep{1981A&A...102...17P}, capping further accretion. The material leaves the system as boosted fast winds, taking away the angular momentum of the accretor star and the orbit. The model is a physically motivated implementation in \texttt{MESA}. However, different binary systems imply varying mass accretion efficiencies. For example, sdB and main-sequence binaries do not show evidence of substantial mass accretion \citep[e.g.][]{2017A&A...605A.109V} but some sdOB and Be star binaries indicate a high accretion efficiency \citep[e.g.][]{2024ApJ...962...70K}. Further theoretical and observational research is needed to better understand how and under what conditions mass transfer can be efficient.

\begin{figure*}\center
\includegraphics[width=\textwidth]{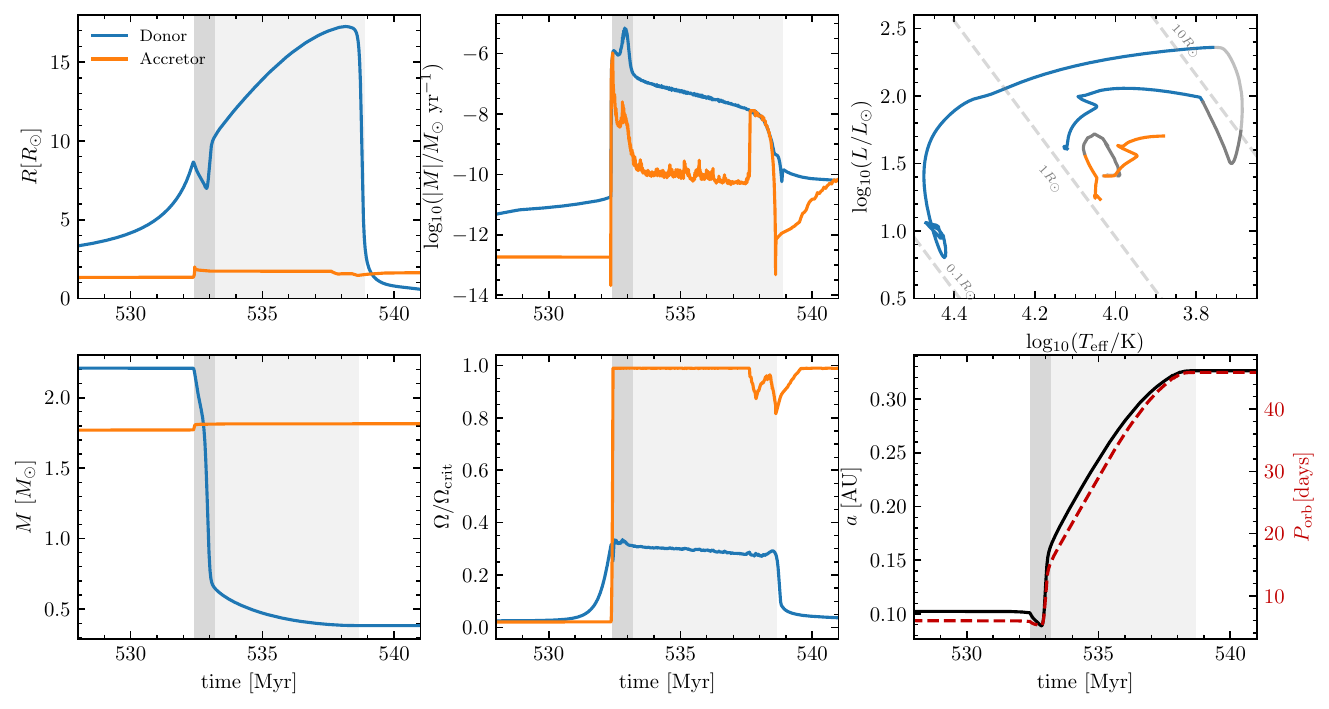}\\
\caption{Evolution of the binary system properties during the mass transfer phase. The different panels show the binary stars' radii (top left), the absolute values of mass change rates (top middle), the Hertzsprung–Russell (HR) diagram for both stars (top right), the binary stars' masses (bottom left), the surface rotation velocities over their critical values (bottom middle), and the orbital separation and period (bottom right) as a function of time. The blue and orange lines indicate the donor and accretor stars, respectively. The HR diagram shows the fast and slow mass transfer phases with dark grey and light grey lines, respectively. The other plots are marked with dark grey and light grey shaded areas.}
\label{fig:b1}
\end{figure*}

Figure \ref{fig:b1} top panels show the evolution of the radii, of the absolute values of mass change rates, and of the position in the Hertzsprung–Russell (HR) diagram for both stars in our reference binary model. Figure \ref{fig:b1} bottom panels show the evolution of the component masses, of the surface angular velocities over their critical values, and of the orbital separation $a$ and period $P_{\rm{orb}}$. After leaving the main sequence, the donor star ignites hydrogen in a shell around the helium core. It expands rapidly, filling the Roche lobe at around $532.3\,\rm{Myr}$, initiating a fast mass transfer phase for about $0.8\,\rm{Myr}$. The mass transfer rate reaches the maximum of $\sim 10^{-5}\,M_{\odot}\,\rm{yr^{-1}}$ at about $532.9\,\rm{Myr}$. The donor star loses about $1.5\,M_{\odot}$ during the fast mass transfer phase. Then, the binary enters a slow mass transfer phase with a mass transfer rate of $\sim 10^{-8}-10^{-7}\,M_{\odot}\,\rm{yr^{-1}}$, lasting about $6\,\rm{Myr}$. At the beginning of the mass transfer phase, the accretor accepts all the material from the donor. In a short period of time the accretor is spun up, then the accretion rate drops quickly. After the mass transfer process, the donor's hydrogen envelope is stripped, and it shrinks significantly, becoming a subdwarf. Although it is known that subdwarfs are substantially affected by atomic diffusion and gravitational settling, which causes them to show almost a hydrogen-pure atmosphere quickly, we do not account for that detail here since we focus on the mass transfer phase. The mass of the donor star decreases from $2.21\ M_{\odot}$ to $\simeq 0.38\ M_{\odot}$ 
and the accretor accretes $\approx 0.04\ M_{\odot}$. Because angular momentum is lost from the system during mass transfer, the binary orbit widens from $0.10\,\rm{AU}$ to $0.33\,\rm{AU}$, meaning the orbital period increases from $6$ days to $46$ days.

\subsection{Coupling of MESA and REBOUND}
\label{sec2.2}

To build up circumbinary planet systems in \texttt{REBOUND}, we first add two stars with properties matching those of the \texttt{MESA} binary at the starting point of tracing the planet's dynamical evolution. Then, we add a planet in the simulation that can be described by its mass, radius, and orbital elements. We use the \texttt{WHFast} integrator, which is a second-order symplectic Wisdom Holman integrator \citep{2015MNRAS.452..376R}, with a fixed timestep of $10^{-3}\,\rm{yr}$ to calculate the dynamical evolution of the planet. 

Throughout the calculation process, we treat the evolution of the central binary as an isolated binary, which is precalculated with \texttt{MESA}. As a result, it is important to refresh and synchronize the binary parameters within \texttt{REBOUND} properly. We linearly interpolate all the binary properties as a function of time to generate a new set of \texttt{MESA} binary output, similar to what was achieved by \citet{2022MNRAS.510.6001B}. However, updating the binary properties during binary evolution requires more thorough scrutiny. In cases where the binary exists in a stable state, for example, the long-lasting main-sequence evolution, it is feasible to update the binary properties at a low frequency.

In contrast, when the binary is experiencing dramatic changes, such as during a rapid mass transfer phase, a reduction in the time interval for the updates becomes essential. In \texttt{MESA}, adaptive timesteps are controlled by various factors, generally decreasing when the system undergoes significant changes. Therefore, it is natural to use the time series from \texttt{MESA}’s output as the time points to update the binary properties. However, this approach is insufficient because the minimum timestep required for the binary evolution is longer than that for the planet's dynamical evolution. This discrepancy would introduce systematic errors in the cases where the binary state changes substantially within a single \texttt{MESA} timestep. 

Thus, we introduce an input parameter defined as the change of a quantity within one \texttt{MESA} timestep to further adjust the timesteps for the updates. During mass transfer, one of the most rapidly changing parameter is the donor star mass. As a result, we monitor the change of the donor star mass $\Delta M_{1} = M_{1,k+1} - M_{1,k}$ for the mass transfer phase, where $k$ denotes the step for \texttt{MESA} output. If $\Delta M_{1}$ is larger than a specific threshold, we split this particular step evenly into a greater number of smaller intervals to ensure that $\Delta M_{1}$ for a single new timestep is below the threshold. In this way, we generate a revised sequence of binary properties, predetermined by the interpolated \texttt{MESA} binary output, along with re-calibrated time intervals, in preparation for the subsequent computation of planetary dynamics.

\begin{figure}\centering
\includegraphics[width=0.48\textwidth]{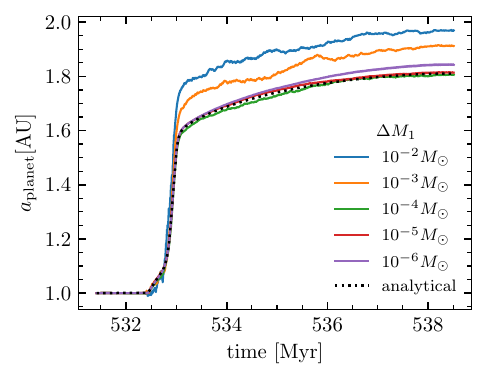}\\
\caption{Evolution of the semi-major axis of the planet with different thresholds for the donor star mass change ($\Delta M_{1}$) within a single customized timestep. The dotted line indicates the analytical orbital evolution due to mass loss from the central object.
}
\label{fig:c1}
\end{figure}

In order to obtain an appropriate threshold for $\Delta M_{1}$, a convergence test is conducted by exploring different limits of $\Delta M_{1}$. We consider a simplified model where the only circumbinary planet is a test particle with an initial separation of $1\,\rm{AU}$ from the center of the mass in a circular orbit. We adopt five thresholds for $\Delta M_{1}$, ranging from $10^{-2}\,M_{\odot}$ to $10^{-6}\,M_{\odot}$, spaced apart by one order of magnitude. We calculate the orbital evolution of the planet from about $2\,\rm{Myr}$ prior to the mass transfer phase until the end of the mass transfer phase. Figure \ref{fig:c1} shows the evolution of the semi-major axis of the testing planet $a_{\rm{planet}}$ under different thresholds for $\Delta M_{1}$ through the binary mass transfer phase. Above $10^{-3}\,M_{\odot}$, we find unexpected fluctuations and large deviations from other tracks, indicating large errors. The values of $10^{-4}\,M_{\odot}$ and $10^{-5}\,M_{\odot}$ lead to similar final $a_{\rm{planet}}$. However, with a further reduction to $10^{-6}\,M_{\odot}$, the evolutionary track diverges, deviating from convergence. 

To verify our calculation and to find out a suitable threshold for $\Delta M_{1}$, we do an analytical calculation for the planet's orbit. In our testing case, the distance between the planet and the binary exceeds the binary separation by a considerable degree (initially $a_{\rm{planet}}/a \sim 10$), which enables us to treat the binary as a single object. Consequently, the mass loss resulting from the binary mass transfer process can be seen as mass loss from a single system. Then, assuming no change in the planet's mass, the change in the orbital separation between the planet and the binary is
\begin{equation}\label{af}
a_{f} = a_{i}\frac{M_{1,i}+M_{2,i}}{M_{1,f}+M_{2,f}},
\end{equation}
where $a_{i}$ and $a_{f}$ represent the orbital separations before and after mass transfer, respectively, while $M_{1,f}$ and $M_{2,f}$ denote the masses of the stars after mass transfer.

To do the comparison, we calculate analytically the expected evolution of the semi-major axis of the planet by inputting the binary masses in equation \ref{af} at each timestep to update binary properties. The black dotted line in Figure \ref{fig:c1} shows the semi-major axis evolution of the planets because of the mass loss of the central object. The analytical calculation aligns most closely with the case of $10^{-5}\,M_{\odot}$, resulting in a comparable final $a_{\rm{planet}}$. In the case of $10^{-6}\,M_{\odot}$, the newly determined time intervals for \texttt{MESA} seem too small to allow the integrator to stabilize the planet's orbit. The errors accumulate through the too-frequent updates of the binary properties, leading to an excess of $a_{\rm{planet}}$. As a result, we adopt $10^{-5}\,M_{\odot}$ as the threshold for $\Delta M_{1}$ through the mass transfer phase in our calculation.



\section{Evolution of a Single Circumbinary Planet}
\label{sec3}

With the binary evolution integrated in \texttt{REBOUND}, we demonstrate its use by modeling the dynamical evolution of a single circumbinary planet through the mass transfer phase.

\begin{figure}\centering
\includegraphics[width=0.48\textwidth]{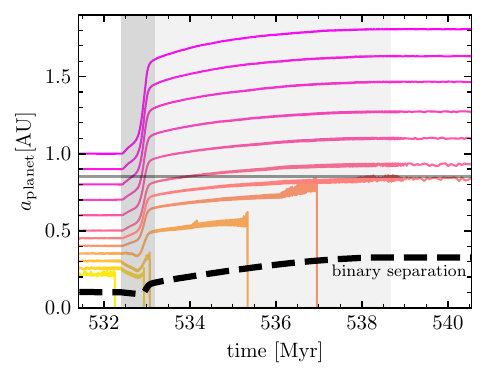}\\
\caption{Evolution of the semi-major axis for Jupiter-like planets with different initial separations. The black dashed line represents the binary separation. The horizontal grey line indicates the closest stable orbit for the circumbinary planet after mass transfer. The slow and fast mass transfer phases are marked with dark grey and light grey shaded areas, respectively.
}
\label{fig:c2}
\end{figure}

We consider a Jupiter-like planet ($1\,M_{\rm{Jup}}$ and $1\,R_{\rm{Jup}}$) in a circular orbit starting from $1\,\rm{Myr}$ prior to the onset of mass transfer and progressing through the mass transfer phase. The initial separations for the planet and the center of mass are from $0.2\,\rm{AU}$ to $0.5\,\rm{AU}$, spacing apart by $0.05\,\rm{AU}$, and from $0.5\,\rm{AU}$ to $1.0\,\rm{AU}$ with a step size of $0.1\,\rm{AU}$. Figure \ref{fig:c2} shows the evolution of $a_{\rm{planet}}$ for Jupiter-like planets with different initial separations. The black dashed line indicates the separation of the central binary stars. We can see that the planets with initial $a_{\rm{planet}}$ below about $0.3\,\rm{AU}$ are quickly driven to an unstable interaction with the binary by the gravitational forces of the central binary. In the case of $0.35\,\rm{AU}$, the planet exhibits instability and migrates inward during the onset of the mass transfer phase. As the system enters the fast mass transfer phase, the planet's orbit rapidly expands. During the subsequent slow mass transfer phase, the orbit of the planet displays intensified oscillations as the binary separation gradually increases. Eventually, in the midst of the slow mass transfer phase, the planet's orbit becomes highly unstable, which likely leads to engulfment by the binary or ejection from the system. The planet with an initial separation of $0.4\,\rm{AU}$ also survives the fast mass transfer phase and enters the chaotic region in-between the stars in the subsequent slow mass transfer phase due to an escalation in orbital instability. The planets initially separated above $0.4\,\rm{AU}$ survive the whole mass transfer process. They all experience a rapidly accelerating orbital expansion as the mass transfer rate attains its maximum, followed by a decelerated expansion during the subsequent slow mass transfer phase. The closest stable orbit after mass transfer is located at $\approx 0.85\,\rm{AU}$, which is $\approx 2.5$ times the binary separation after that phase. As the planets are farther away from the central binary, the oscillations of the orbit gradually diminish in intensity in the late slow mass transfer phase. Interestingly, for circumbinary planets around main-sequence binaries, it has been found that the closest stable orbit is located at approximately $2$ to $2.5$ times the binary separation for low eccentricity binaries \citep[e.g.,][] {1986A&A...167..379D,1989A&A...226..335D,1997AJ....113.1445W,1999AJ....117..621H,2002CeMDA..82..143P}.

\section{\texttt{NBSE} Tides}
\label{sec4}

Tidal forces grow stronger as the distance between the star and the planet decreases. In the region where the planet's semi-major axis is not significantly larger than the binary separation, it is essential to consider tidal effects on the planet, which can alter the orbital evolution. We apply the prescription in \citet{2023ApJ...948...41L}, where they implement self-consistent spin, tidal, and dynamical equations of motion in the \texttt{REBOUNDx} framework. The tidal prescription is based on the approach in \citet{1998ApJ...499..853E}, considering the acceleration from the quadrupolar distortion:

\begin{equation}\label{tide1}
\boldsymbol{f}_{\rm QD,1} = r_{1}^5k_{\rm L,1}(1+\frac{m_{2}}{m_{1}}) \cdot [\frac{5(\boldsymbol{\Omega}_{1} \cdot \boldsymbol{d})^2\boldsymbol{d}}{2d^7} - \frac{\Omega_{1}^2\boldsymbol{d}}{2d^5} - \frac{(\boldsymbol{\Omega}_{1} \cdot \boldsymbol{d})\boldsymbol{\Omega}_{1}}{d^5} - \frac{6Gm_{2}\boldsymbol{d}}{d^8}] ,
\end{equation}
and the acceleration from tidal damping:

\begin{equation}\label{tide2}
\boldsymbol{f}_{\rm TF,1} = -\frac{9\sigma_{1} k_{\rm L,1}^2 r_{1}^{10}}{2d^{10}}(m_{2} + \frac{m_{2}^2}{m_{1}}) \cdot [3\boldsymbol{d}(\boldsymbol{d} \cdot \dot{\boldsymbol{d}})+(\boldsymbol{d}\times \dot{\boldsymbol{d}} - \boldsymbol{\Omega}_{1}d^2)\times \boldsymbol{d}] ,
\end{equation}
where $r_{1}$ is the radius of object 1, $k_{\mathrm{L},1}$ denotes the Love number of object 1 , while $m_{1}$ and $m_{2}$ are the masses of object 1 and object 2, respectively. $\Omega_{1}$ represents the angular velocity of object 1, assuming uniform rotation, and $\sigma_{1}$ is the dissipation constant of object 1. The parameter $d$ denotes the distance between the two objects, and $G$ is the gravitational constant. 

In the case of the binary stars experiencing a mass transfer process, the properties of the stars change significantly, especially for the donor star. As a result, it is imperative to obtain accurate parameters for the equation above at different stages. For the stars, we have the stellar profiles provided by \texttt{MESA}, allowing us to calculate all the parameters self-consistently. The Love number is two times the apsidal motion constant $k$, which can be calculated with the relation \citep{1939MNRAS..99..451S}:
\begin{equation}\label{k1}
k = \frac{3-\eta_{2}}{4+2\eta_{2}} .
\end{equation}
$\eta_{2}$ is a function of the radius $r$ that can be obtained from the equation \citep{1939MNRAS..99..451S}:
\begin{equation}\label{k2}
r\frac{d\eta_{2}}{dr} + \eta_{2}(1-\eta_{2}) + 6\frac{\rho}{\overline{\rho}}(\eta_{2}+1) - 6 = 0  ,
\end{equation}
where $\rho$ is the density at $r$ and $\overline{\rho}$ is the mean density interior to $r$. We save the density profiles of the stars every ten steps in \texttt{MESA} to calculate $\eta_{2}$ and then the Love numbers. Afterward, we perform linear interpolation over the time series to determine the evolution of the Love number for both stars. As for the dissipation constant, it is connected with the Love number and the lag time $\tau$ \citep{2023ApJ...948...41L}:
\begin{equation}\label{k3}
\sigma_{1} = \frac{3r_{1}^5k_{L,1}}{4G\tau_{1}}.
\end{equation}
The lag time is related to the typical tidal timescale $T$, defined in \citet{1981A&A....99..126H}:
\begin{equation}\label{k4}
T_{1} = \frac{r_{1}^3}{Gm_{1}\tau_{1}}.
\end{equation}
Then, we follow the same method in \texttt{POSYDON} configuration to calculate the quantity $k/T$ \citep[see Sec.4.1 in ][]{2023ApJS..264...45F} to get access to all the parameters involved in the calculation of tides for the stars. For the Jupiter-like planet, we adopt a typical $k_{L}$ of $0.565$. As for the dissipation constant $\sigma$, we use the simplified assumption $Q^{-1} \sim 2n \tau$ \citep{2023ApJ...948...41L} and set $Q = 10^{4}$ to calculate $\tau$ and hence $\sigma$, where $Q$ is the specific dissipation function \citep{1963MNRAS.126..257G} and $n$ is the orbital mean motion. 

\begin{figure}\centering
\includegraphics[width=0.48\textwidth]{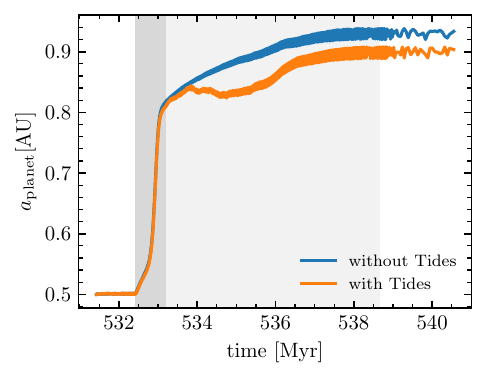}\\
\caption{Evolution of the semi-major axis for a Jupiter-like planet, initially situated at a separation $0.5\,\rm{AU}$, both with and without the inclusion of tidal effects. The slow and fast mass transfer phases are marked with dark grey and light grey shaded areas, respectively.
}
\label{fig:tides}
\end{figure}

The tidal forces between the planet and two stars are performed separately. We update the stellar properties involved in calculating tidal effects with the newly generated \texttt{MESA} time series. We ignore the tidal influence of the planet on the stars' spins as the effects between the stars themselves are predominant, which are accounted for in the \texttt{MESA} simulation. In Figure \ref{fig:tides}, we show the evolution of the semi-major axis of a Jupiter-like planet with an initial $a_{\rm{planet}}$ of $0.5\,\rm{AU}$, both with and without accounting for tides during the mass transfer phase. At the beginning of the simulation, although the planet is close to the stars, the Love number, dissipation constant, and radius of the stars are at a low level, leading to a negligible impact on the planet's orbit. Then, the donor star keeps expanding, and the Love number increases to a high level. During the fast mass transfer phase, the mass loss from the binary system dominates the dynamical evolution of the planet. After entering the slow mass transfer phase, despite the Love number initiating a decline, the donor star continues to expand, amplifying the significance of tidal effects within the system. Then, as the donor gets stripped and the separation increases, the tidal effect becomes less pronounced. After the mass transfer phase, tidal effect becomes negligible again because the donor star is fully stripped. Our results highlight that continuously tracking the stellar properties for adjusting the parameters used in calculating tidal effects is indispensable when stars undergo rapid changes, such as during mass transfer.

\section{Discussion and Summary}
\label{sec5}

In this work, we developed \texttt{NBSE}, a framework designed to incorporate binary evolution data from the stellar evolution code \texttt{MESA} into the N-body simulation code \texttt{REBOUND}. It thus enables studies of the dynamical evolution of circumbinary planets, even throughout phases of binary interaction. To demonstrate how \texttt{NBSE} can be used, we constructed a reference binary model with initial masses of 2.21$M_\odot$ and 1.76$M_\odot$ (corresponding to an initial mass ratio of 0.8) and an orbital period of 6 days, corresponding to an initial orbital separation of 0.1 AU. The more massive star initiates stable mass transfer during the hydrogen-shell burning stage prior to the red giant branch (the Hertzsprung gap), which results in complete envelope loss and the formation of a $0.38\,M_{\odot}$ subdwarf in a wide orbit with a period of 46 days, corresponding to a binary separation of 0.33 AU. 
In post-processing, we then adopt the orbital separation, stellar masses and radii from the binary evolutionary model, and input these properties into an N-body simulation, where we represent both stars as spherical bodies with changing orbital phase, and we also integrate the orbit of a surrounding, coplanar, 1$M_{\rm Jup}$ circumbinary planet. 

To mitigate the systematic errors originating from altering binary parameters during the computation of planetary dynamical evolution, we introduce a method for re-calibrating the time sequence for updating \texttt{MESA} binary properties in \texttt{REBOUND} (see Sect.~\ref{sec2.2}). We consider a single Jupiter-like planet around the binary and calculate its dynamical evolution through the mass transfer phase. In our reference model, we find that the nearest stable orbital separation of the circumbinary planet is $\approx 2.5$ times the binary separation after the mass transfer phase, which corresponds to $\approx 4.5$ times the initial separation before mass transfer. The mass accretion efficiency within the central binary system can influence the closest stable orbit. If the mass transfer is more conservative, less mass is lost from the binary, and the expansion of the planet's orbit should be less significant. In this case, the stabilizing zone for the circumbinary planet can be smaller. Similarly, if the mass ratio of the initial binary were more extreme, the system may tighten, leaving even shorter orbital separations for the planet dynamically stable. 

To include tidal forces acting on the planet, we apply the implementation of \citet{2023ApJ...948...41L} in the extended library \texttt{REBOUNDx}, adopting adaptive parameters for tidal effects based on the structure of the stars. We found that the mass loss during the rapid mass transfer phase dominates the planet's dynamical evolution, making the effect of tides negligible. After the rapid mass transfer phase, the significance of tidal effects highly depends on the stellar structure, which undergoes substantial changes during the mass transfer phase. In the case of our simulation, tides become evident at the beginning of the slow mass transfer phase and fade away as the donor star gets stripped and the separation increases.

Our reference binary forms a sdB and A-type star binary with an orbital separation of $\sim 0.33\,\rm{AU}$. If a planet survives the mass transfer phase, we expect to find it in an orbit wider than $\sim 0.85\,\rm{AU}$ around such a binary after the envelope-stripping is complete. 
Although orbital oscillations are present in the planetary orbit after the host binary has interacted, we expect that the configuration reaches dynamical stability in our model since tides are weak and the gravitational perturbation remains constant. This means that the circumbinary planet likely remains on a similar orbit until it is perturbed again (for example, by passing stars or the stellar evolution of the central binary).  
The long-term dynamical evolution of the circumbinary planet can be further explored with \texttt{NBSE}, including additional physics such as tidal decay \citep{2018AJ....156...52S}. 

We expect a similar formation path to produce binaries with sdB orbiting companion stars down to K-type stars ($\sim$0.7$M_\odot$), with the companion mass range determined by varied binary physics, such as how stable mass transfer is, the mass transfer efficiency and angular momentum loss \citep{1997A&A...327..620S}. Similarly, it could be that subdwarfs orbiting white dwarfs potentially retain circumbinary planets.  
As more and more subdwarfs in binary systems are discovered, it is intriguing to search for circumbinary planets around them. Furthermore, subdwarfs are very hot ($\sim 25,000\rm{K}$) and $\sim 10$ times more luminous than the Sun. It is, therefore, possible that the effects of radiation on the planets are important. The real survival region and habitable zone for these planets are subject to these effects.

In our reference model, the highest mass transfer rate is $\sim 10^{-5}\,M_{\odot}\,\rm{yr^{-1}}$. With the threshold $\Delta M_{1} = 10^{-5}\,M_{\odot}$, the minimal timestep for \texttt{MESA} is about $1\,\rm{yr}$, which is still much longer than the timestep of the integrator for the dynamical evolution. As a result, the updates of binary data would not lead to a loss of accuracy as long as the changes are adiabatic. If the change is very rapid, like CE evolution, a finer time resolution or a more suitable integrator is required. For a different binary model, a new convergence test must be conducted to determine the threshold of the changing parameter. 

Furthermore, we have ignored the interaction between the planet and the material that is lost from the binary. For high mass-loss rates, the gas density of the lost material is likely also higher, suggesting that the influence could be the strongest during the rapid phase at the beginning of mass transfer (see the second panel of Figure~\ref{fig:b1}). While estimating the impact of the ejected material on the planet would be interesting for an isotropic outflow, it is possible that the ejecta is not isotropic. If, for example, the material is ejected through jets, its influence on a circumbinary planet would be negligible. It is also worth noting that subdwarfs stripped through successful CE ejection would produce an ejecta that is more dangerous to circumbinary planets since it is denser.  
To carefully account for the influence of the eject is an interesting next step in our investigation, but beyond the scope of this study.

Our model is a first step towards understanding the dynamical effect on planets that orbit interacting binaries and is therefore approximate. However, we can already note an interesting consequence that binary interaction has on planetary stability: in our example system, a planet is allowed to orbit at about 1 AU from the central host star when the system evolves beyond the main-sequence evolution. This would not have been possible if the host star were single since it then would have engulfed the planet, swallowing it whole at red giant branch or asymptotic giant branch \citep{2007ApJ...661.1192V,2008MNRAS.386..155S}. In this regard, the binary interaction preserves tight-orbit planets. 

We created \texttt{NBSE} as a starting point for investigating in greater detail the planetary architecture and the conditions of disruption, pollution, and engulfment for planets around single stars and, in particular, interacting binaries. Further development and extension of \texttt{NBSE} could potentially enable us to explore the dynamical evolution of planets with high precision and accuracy around binary systems undergoing more dramatic processes, such as CE evolution and stellar mergers.

\section*{Acknowledgments}

We thank the participants of the 2023 Kavli Summer Program in Astrophysics, hosted by the Max Planck Institute for Astrophysics and funded by the Kavli Foundation. In particular, Holly Preece, Selma de Mink, and Stephen Justham for their feedback and comments on our work. ZX acknowledges support from the China Scholarship Council (CSC). ST acknowledges the funding from the European Union’s Horizon 2020 research and innovation program under the Marie Sk\l{}odowska-Curie grant agreement No 101034413. AAT acknowledges support from the Horizon Europe research and innovation programs under the Marie Sk\l{}odowska-Curie grant agreement no. 101103134.

\section*{Data Availability}

The scripts used to generate the data for this work can be requested at Zepei.Xing@unige.ch and Santiago.Torres@ist.ac.at. The final scripts will be made publicly available upon publication.


\bibliographystyle{mnras}
\bibliography{Zepei_Torres_etal} 




\bsp	
\label{lastpage}
\end{document}